%% file: main.tex
\documentclass[10pt,twocolumn,letterpaper]{article}

\usepackage{iccv}              


\definecolor{iccvblue}{rgb}{0.21,0.49,0.74}
\usepackage[accsupp]{axessibility}
\usepackage[pagebackref,breaklinks,colorlinks,allcolors=iccvblue]{hyperref}
\usepackage{multirow}
\usepackage{booktabs}
\usepackage{arydshln}
\usepackage{enumitem}
\usepackage{algorithmicx}
\usepackage{algorithm}
\usepackage{algpseudocode}

\def\paperID{6500} 
\def\confName{ICCV}
\def\confYear{2025}

\title{Discretized Gaussian Representation for Tomographic Reconstruction}

\author{
    Shaokai Wu\textsuperscript{\rm 1}, 
    Yuxiang Lu\textsuperscript{\rm 1}, 
    Yapan Guo\textsuperscript{\rm 4},
    Wei Ji\textsuperscript{\rm 2}, 
    Suizhi Huang\textsuperscript{\rm 1}, 
    Fengyu Yang\textsuperscript{\rm 3}, \\
    Shalayiding Sirejiding\textsuperscript{\rm 1}, 
    Qichen He\textsuperscript{\rm 1}, 
    Jing Tong\textsuperscript{\rm 1}, 
    Yanbiao Ji\textsuperscript{\rm 1}, 
    Yue Ding\textsuperscript{\rm 1}, 
    Hongtao Lu\textsuperscript{\rm 1*} \\
    \textsuperscript{\rm 1}Shanghai Jiao Tong University, 
    \textsuperscript{\rm 2}National University of Singapore, \\
    \textsuperscript{\rm 3}Yale University,
    \textsuperscript{\rm 4}Suzhou Xiangcheng People's Hospital \\
    {\tt\small \{shaokai.wu, luyuxiang\_2018, huangsuizhi, salaydin\}@sjtu.edu.cn}\\
    \tt \small \{ayombeach, tj\_19\_hf, jiyanbiao, dingyue, htlu\}@sjtu.edu.cn\\
    \tt \small guo\_yapan@outlook.com, jiwei@nus.edu.sg, fengyu.yang@yale.edu
    }
\begin{document}
\maketitle
\begin{abstract}
Computed Tomography (CT) enables detailed cross-sectional imaging but continues to face challenges in balancing reconstruction quality and computational efficiency. While deep learning-based methods have significantly improved image quality and noise reduction, they typically require large-scale training data and intensive computation. Recent advances in scene reconstruction, such as Neural Radiance Fields and 3D Gaussian Splatting, offer alternative perspectives but are not well-suited for direct volumetric CT reconstruction. In this work, we propose Discretized Gaussian Representation (DGR), a novel framework that reconstructs the 3D volume directly using a set of discretized Gaussian functions in an end-to-end manner. To further enhance efficiency, we introduce Fast Volume Reconstruction, a highly parallelized technique that aggregates Gaussian contributions into the voxel grid with minimal overhead. Extensive experiments on both real-world and synthetic datasets demonstrate that DGR achieves superior reconstruction quality and runtime performance across various CT reconstruction scenarios. Our code is publicly available at \href{https://github.com/wskingdom/DGR}{https://github.com/wskingdom/DGR}.

\end{abstract}
\let\thefootnote\relax    
\footnotetext[0]{\textsuperscript{*}Corresponding author.}

\section{Introduction}
\begin{figure}[t]
  \centering
  \setlength{\belowcaptionskip}{-3mm}
  \includegraphics[width=.985\linewidth]{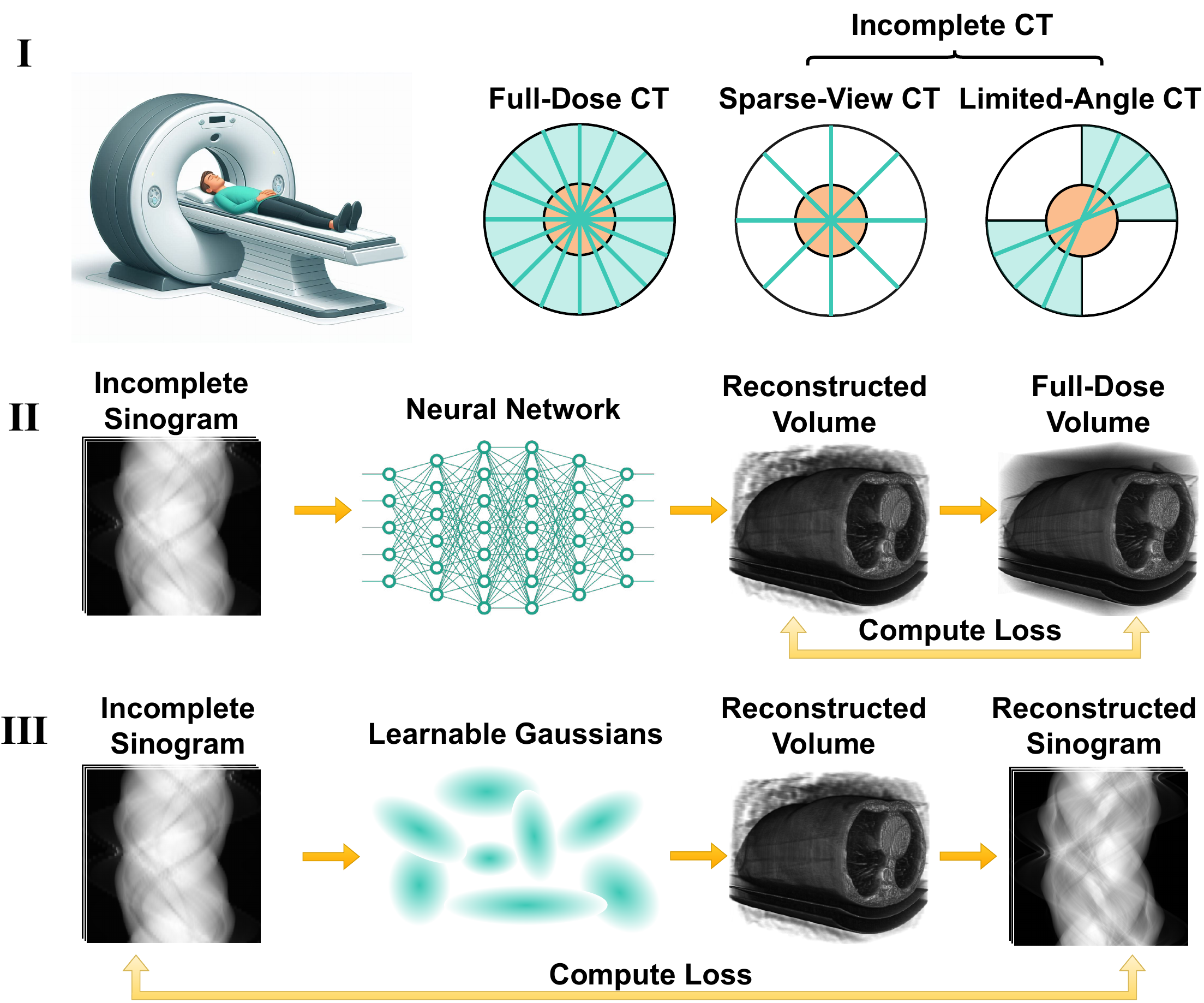}
\caption{\uppercase\expandafter{\romannumeral1}: CT imaging workflow: X-ray projections are acquired from multiple angles.  
\uppercase\expandafter{\romannumeral2}: Deep Learning-based Reconstruction (DLR): Networks are trained on paired projection-image datasets to reconstruct volumes, requiring extensive pre-training.  
\uppercase\expandafter{\romannumeral3}: Instance reconstruction (exemplified by 3DGS): Instance-adaptive optimization of 3D Gaussians via differentiable rendering, tailored for each instance without the need for training datasets.}
  \label{teaser}
\end{figure}

When a patient undergoes a Computed Tomography (CT) scan at a hospital, the initial output is not the familiar image seen on the screen, but rather a series of projection data generated by X-ray detectors~\cite{CT-Survey-2023}. This data is then processed by a CT reconstruction algorithm to transform the raw information into a 3D volume, which radiologists examine to assess the patient’s condition. This task is challenging due to three key factors. Firstly, the ionizing radiation exposure from a complete CT scan poses significant health risks to patients, limiting the projections that can be acquired~\cite{r2-gaussian}. Secondly, time constraints in emergency situations require fast reconstruction to enable timely diagnosis~\cite{x-gaussian}. Thirdly, the method must be generalizable across various CT configurations, such as different scan geometries, to accommodate heterogeneous imaging protocols and patient populations.

As illustrated in Figure~\ref{teaser}, existing CT reconstruction methods broadly fall into two categories: Deep Learning-based Reconstruction (DLR) and instance reconstruction. Early DLR approaches, such as FBPConvNet~\cite{FBPConvNet} and RED-CNN~\cite{REDCNN}, primarily focused on image denoising within the reconstructed domain. Subsequent methods like HDNet~\cite{HDNet} advanced to joint optimization across both projection and image domains. More recently, advanced diffusion-based DLR methods, including DiffusionMBIR~\cite{DiffusionMBIR} and SWORD~\cite{SWORD}, leverage score-based diffusion models to guide the reconstruction process. Despite these advancements, DLR methods face persistent challenges: high training costs and limited generalization across diverse CT scanner configurations. For instance, models trained on Cone-Beam CT data often perform poorly on Fan-Beam CT systems, and networks optimized for chest scans may show degraded performance on head scans.

In response to these challenges, instance reconstruction methods have emerged, specifically designed to yield patient-specific results. Inspired by the success of scene reconstruction techniques, particularly Neural Radiance Fields (NeRF)~\cite{NeRF} and 3D Gaussian Splatting (3DGS)~\cite{3DGS}, numerous adaptations have been proposed for CT reconstruction. Exemplars of NeRF-based methodologies, such as NAF~\cite{naf}, Intra-Tomo~\cite{intratomo}, and SAX-NeRF~\cite{sax}, implicitly represent the tomographic scene as a continuous function of 3D spatial coordinates and optimize a neural network within the projection domain. However, these methods typically require hours per reconstruction, which renders them impractical for real-time clinical deployment.
On the other hand, 3DGS-based methodologies, such as 3DGR-CAR~\cite{car} and X-Gaussian~\cite{x-gaussian}, leverage explicit Gaussian functions to represent tomographic scenes. Nevertheless, directly applying 3DGS to CT leads to a strong integration bias~\cite{r2-gaussian} due to density inconsistency inherent in 3DGS. While R$^2$-Gaussian~\cite{r2-gaussian} attempts to mitigate these issues with a rectified 3DGS rasterizer and voxelizer, it introduces computational overhead stemming from the fundamental discrepancy between 3DGS's view-oriented rendering and the requirements of discretized tomographic reconstruction.

These limitations highlight a critical gap: existing DLR methods are hindered by dataset dependency and training complexity, while instance reconstruction paradigms are not well-aligned with tomographic objectives. Motivated by these challenges, we rethink the framework's design based on three core principles: discretized representation, efficient reconstruction, and unified global optimization.

From these foundational considerations, we propose our Discretized Gaussian Representation (DGR) method.
For \textbf{representation}, given that the target volumetric grid is discretized, we directly model the 3D volume as a set of discretized Gaussian functions. To ensure differentiability and accurate preservation of continuous Gaussian contributions within local grid regions, we developed a novel alignment technique.
For \textbf{reconstruction}, we advance beyond traditional voxelization with our highly parallelized Fast Volume Reconstruction method, which can reconstruct a \textbf{$256\times 256\times 256$} volume from over 150,000 Gaussians in just 0.09 seconds per iteration.
In terms of \textbf{optimization}, the synergy of our lightweight representation and fast reconstruction allows for joint optimization of all Gaussians, a departure from 3DGS-based methods that rely on selective view or block optimization.
We evaluate DGR on three public CT datasets: FIPS~\cite{fips}, AAPM-Mayo LDCT~\cite{mayo}, and FUMPE~\cite{fumpe}, covering both real-world and synthetic data. Comprehensive experiments across multiple reconstruction tasks demonstrate that DGR surpasses state-of-the-art DLR and instance reconstruction methods in both quantitative metrics and qualitative assessments.

Our main contributions can be summarized as follows:
\begin{itemize}[leftmargin=4mm,noitemsep,topsep=0pt]
\item We propose DGR, an end-to-end CT reconstruction method that directly reconstructs 3D volumes by learning from discretized Gaussian functions.
\item We introduce a fast volume reconstruction technique that efficiently aggregates Gaussian contributions into a discretized volume using a highly parallelized approach, significantly reducing reconstruction time.
\item We demonstrate DGR's superior performance over both DLR and instance reconstruction methods through extensive experiments on real-world and synthetic datasets.
\end{itemize}

\vspace{-1mm}
\section{Background}

\begin{figure*}[t]
  \centering
  \setlength{\abovecaptionskip}{1mm}
  \setlength{\belowcaptionskip}{-6mm}
  \includegraphics[width=.95\textwidth]{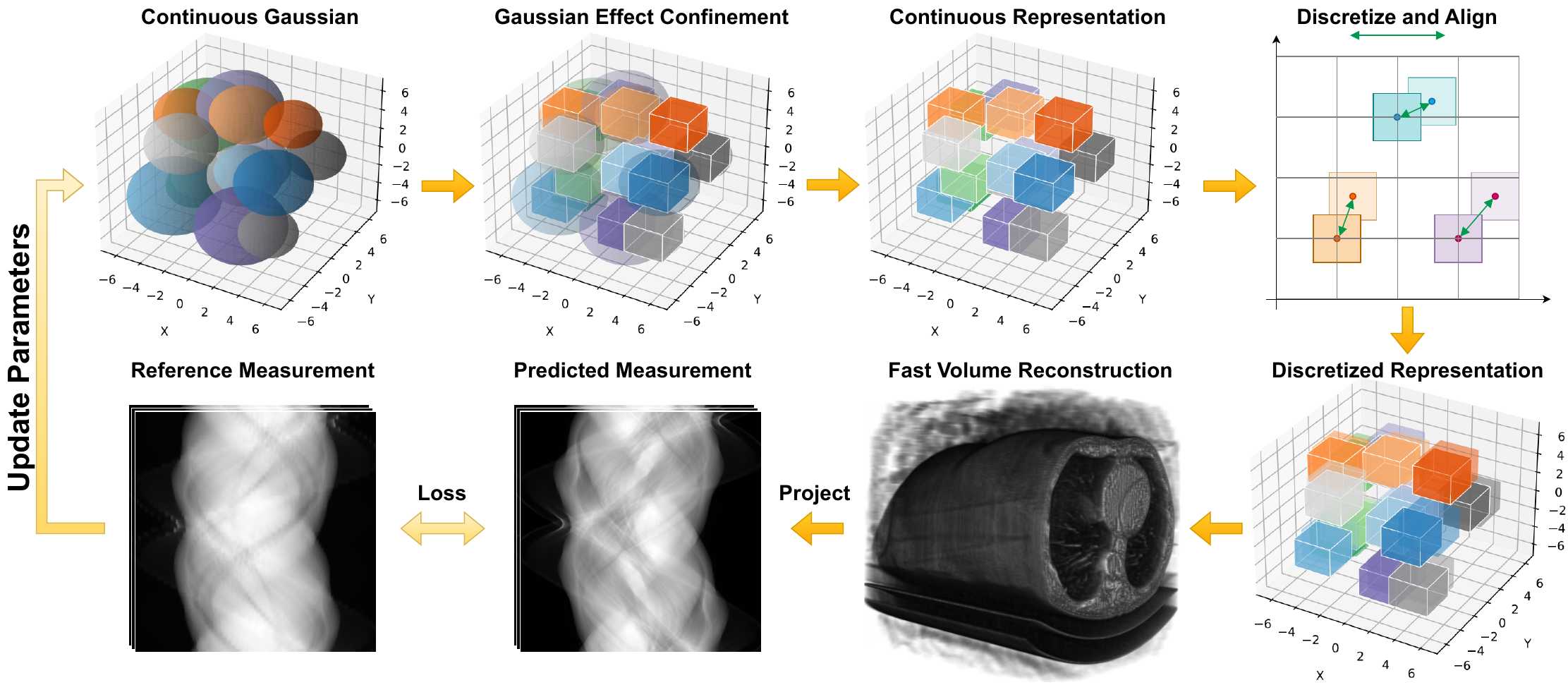}
  \caption{Pipeline of DGR. The 3D volume is initially represented by a set of continuous Gaussians, with each Gaussian's contribution confined to a local region surrounding the voxel. The Gaussians are then discretized onto the 3D grid, where their contributions are aligned to directly reconstruct the entire volume. Fast Volume Reconstruction technique gathers these contributions in a highly parallelized manner. Within each iteration, these Gaussians are reconstructed and then projected into the measurement domain for optimization.}
  \label{fig:overview}
\end{figure*}

\vspace{-1mm}
\subsection{Deep Learning Reconstruction}
\vspace{-1mm}
Deep Learning Reconstruction (DLR) methods have consistently been a prominent area in CT reconstruction since the successful application of deep neural networks in image processing~\cite{unet}. Conventional DLR approaches are primarily categorized into direct-learning and indirect-learning approaches~\cite{PLA}. Direct learning methods involve training networks to directly reconstruct images from projection data (sinograms). Prominent instances include AUTOMAP~\cite{AUTOMAP}, which employs manifold approximation, and iRadonMAP~\cite{he2020Radon}, which utilizes learnable filtering and back-projection techniques. Conversely, indirect learning methods~\cite{FBPConvNet,wavelet1,wavelet2,REDCNN} frame CT reconstruction as a high-level denoising problem. These methods first transform raw projection data into the image domain, typically via filtered back-projection (FBP)~\cite{fbp1967} or iterative reconstruction (IR), and subsequently train networks to map these noisy initial images to their clean counterparts. Certain indirect methods, such as HDNet~\cite{HDNet}, further integrate both projection and image domains' knowledge to guide the reconstruction.

Recent advancements in DLR have led to the development of diffusion models that further guide the reconstruction process. Models like MCG~\cite{MCG}, DiffusionMBIR~\cite{DiffusionMBIR}, and SWORD~\cite{SWORD} learn the score function as a prior, which helps manage noise and artifacts in the reconstructed images. However, diffusion-based methods are computationally more intensive than traditional DLR, limiting their practicality in clinical settings. Despite this, DLR methods can improve generalization and performance by training on large datasets with diverse imaging conditions.

\vspace{-1mm}
\subsection{Instance Reconstruction}
\vspace{-1mm}
Instance Reconstruction-based methods constitute another category of CT reconstruction techniques, focused on reconstructing images through solving an optimization problem. These methods are fundamentally rooted in iterative reconstruction (IR)~\cite{IR}, where an objective function balances data fidelity with regularization terms. These terms impose prior knowledge or constraints, such as sparsity, smoothness, or edge preservation, to guide the reconstruction. Common approaches like algebraic reconstruction techniques (ART)~\cite{herman1973art}, simultaneous iterative reconstruction technique (SIRT)~\cite{sirt}, and total variation (TV)~\cite{tv} regularization are widely adopted to enhance image quality and reduce artifacts compared to direct analytical methods like filtered back projection (FBP)~\cite{fbp1967}.

Recent advancements for CT imaging have seen the adaptation of instance reconstruction methods, such as Neural Radiance Fields (NeRF)~\cite{NeRF} and 3D Gaussian Splatting (3DGS)~\cite{3DGS}. NeRF-based approaches like NAF~\cite{naf} and Intra-Tomo~\cite{intratomo} optimize by minimizing the difference between synthetic and ground-truth projections, incorporating both local and global geometric priors. SAX-NeRF~\cite{sax} further enhances Sparse-View X-ray reconstruction through transformer-based structure-aware modeling. These NeRF-based methods typically require hours of inference time per instance~\cite{r2-gaussian}, making them impractical for clinical use.

Generally, 3DGS-based methods surpass NeRF-based methods in both speed and image quality. Following the original 3DGS, they represent the CT scene as a set of 3D Gaussians. 3DGR-CT~\cite{3dgrct,3dgsct} integrates 3D Gaussians with a differentiable CT projector for Cone-Beam CT reconstruction. 
3DGR-CAR~\cite{car} leverages a U-Net~\cite{unet} to initialize Gaussian centers before employing 3DGS for final reconstruction. DIF-Gaussian~\cite{dif-gaussian} uses 3D Gaussians to model feature distributions for estimating attenuation coefficients. X-Gaussian~\cite{x-gaussian} adapts 3DGS for novel view synthesis in X-ray imaging. However, these 3DGS-based methods are not ideal for direct 3D volume reconstruction. R$^2$-Gaussian~\cite{r2-gaussian} identified integration bias in the standard 3DGS formulation and proposed rectified rasterizer and differentiable voxelizer to mitigate this limitation.

\vspace{-2mm}
\section{Preliminary}
\vspace{-1mm}
\noindent \textbf{Problem Formulation}
Clinical CT imaging involves capturing X-ray projections of the patient from various angles, which are then processed to reconstruct a 3D volume that represents the internal structures. The CT imaging system for a 2D image slice can be described as: 
\begin{equation} 
y = Ax + \beta, 
\end{equation} 
where $x \in \mathbb{R}^{w \times h}$ denotes the 2D image with width $w$ and height $h$, and $y \in \mathbb{R}^{m \times n}$ represents the projection matrix, where $m$ is the number of projection views and $n$ is the number of detector elements per view.
The matrix $A \in \mathbb{R}^{m \times n \times w \times h}$ is the Radon Transform~\cite{radon} matrix, which maps the geometric relationship between the object and the X-ray source/detector system. 
The noise term $\beta \in \mathbb{R}^{m \times n}$ represents measurement noise.

\noindent \textbf{Optimization Goal.}
The goal of CT reconstruction is to accurately estimate the underlying image $x$ from measured projection data $y$. While the Inverse Radon Transform theoretically offers a perfect mapping, real-world issues like data noise and the underdetermined nature of the inverse problem makes perfect reconstruction unattainable.

The reconstruction process is commonly formulated as an optimization problem: 
\begin{equation} 
x^* = \arg\min_x \mathcal{E}(Ax, y) + R(x), 
\end{equation} 
where $\mathcal{E}$ is the data fidelity term, which measures the discrepancy between the estimated measurement $Ax$ and the actual measurement $y$, often using the L1 or L2 norm. The regularization term $R(x)$ enforces properties like smoothness or sparsity in the reconstructed image, with Total Variation (TV)~\cite{tv} regularization being a popular choice~\cite{ADMM-TV}.

\vspace{-1mm}
\section{Methodology}
\vspace{-1mm}
We illustrate our DGR framework in Figure \ref{fig:overview}. 
DGR aims to directly reconstruct the 3D tomographic volume in an end-to-end and efficient manner. We organize our method section into three parts: Discretized Gaussian Representation (Section \ref{sec:DGR}), Fast Volume Reconstruction (Section \ref{sec:recon}), and Global Optimization (Section \ref{sec:optimization}).

\vspace{-1mm}
\subsection{Discretized Gaussian Representation}
\vspace{-1mm}
\noindent \textbf{Continuous Gaussian Representation}
We begin by introducing the basic formulation of the continuous Gaussian representation for tomographic volume reconstruction.
\label{sec:DGR}
We define the Gaussian function centered at $\mu$ with covariance $\Sigma$ as:
\begin{equation}
  \label{eq:basic}
  G(p, \mu, \Sigma) = e^{-\frac{1}{2}(p-\mu)^\top\Sigma^{-1}(p-\mu) },
\end{equation}
where $p \in \mathbb{R}^{d}, d=3$  is the 3D point in the scene (For clarity in the Einstein Summation representation presented in this section, we maintain the notation of a constant $d=3$).  The Gaussian function is bell-shaped and symmetric around the mean $\mu$, with its spread controlled by the standard deviation $\sigma$, determining its extent in 3D space.
We utilize isotropic Gaussians rather than anisotropic ones to represent the 3D volume. This choice is justified by the isotropic attenuation properties of CT tissues, particularly soft tissues, which exhibit minimal directional dependence. As noted in~\cite{kak}, small distributed source elements can be treated as isotropic, consistent with tissue behavior.

\noindent \textbf{Direct Reconstruction}
Let the target volume be denoted by $V \in \mathbb{R}^{w \times h \times c}$, where $w$, $h$, and $c$ correspond to the width, height, and depth (or length) of the volume, respectively. The intensity of a voxel $V(p)$ at a spatial coordinate $p$ within this volume is expressed as the cumulative contribution from all $n$ constituent Gaussians:
\begin{align}
  V(p) &= \sum_{i=1}^{n} G(p,\mu_i,\Sigma_i)\cdot I_i,
  \label{eq:gaussian}
\end{align}
where $I_i \in \mathcal{R}^{+}$ represents the scalar intensity value of the $i$-th Gaussian, functioning as both its individual contribution to the voxel intensity and its weight.  The quadratic form $(p-\mu)^\top\Sigma^{-1}(p-\mu)$ is recognized as the squared Mahalanobis distance, which quantifies the statistical distance of point $p$ from the Gaussian center $\mu$, normalized by the distribution's shape. This quantity will be referred to as $D^2$ in subsequent discussions for conciseness.

The direct computation of every Gaussian's contribution to the entire volume presents a significant challenge, leading to unacceptable computational expense (see Table~\ref{tab:space_time}). For a deeper understanding of this computational complexity, please refer to Appendix \textcolor{iccvblue}{E.2}. To mitigate this issue, we restrict each Gaussian's effect to a localized region around its assigned voxel, as detailed below.

\noindent \textbf{Gaussian Effect Confinement}
The influence of a Gaussian on a given voxel is known to attenuate with increasing distance from the Gaussian's centroid. Leveraging this principle, restricting the contributions to only those Gaussians within a specified proximity of each voxel can substantially accelerate the reconstruction process. 

To formalize this concept, we define the effective proximity for each Gaussian as a cuboid $w_0 \times h_0 \times c_0$, where $w_0$, $h_0$, $c_0$ are all odd integers. This box is centered on the voxel, and we denote the set of all 3D coordinates within this cuboid as $B_0 \subset \mathbb{R}^3$.

Naturally, the specific coordinates within the cuboid $B_0$ depend on the Gaussian function's center $\mu$. For a Gaussian centered at $\mu$, its influence is evaluated at coordinates $B = B_0 - \mu$, and $B$ is a constant tensor. This is because the coordinates within $B_0$ are relative to $\mu$, meaning that subtracting $\mu$ yields a fixed pattern of offsets independent of its absolute position. Mathematically, this reflects the translation invariance of the Gaussian function's shape: shifting $\mu$ only changes the peak's location, not the function's overall form.

Now, we can compute the squared Mahalanobis distance $D^2$ for a given relative coordinate in $B$ with respect to the Gaussian's mean (which is implicitly at the origin for these relative coordinates) as:
\begin{equation}
  D^2 = B^\top C^{-1} B.
\end{equation}
Here, $C$ denotes the covariance matrix of the Gaussian. The contribution of the $i$-th Gaussian to the voxel intensity at a relative coordinate described by $B$ is then given by:
\begin{equation}
  e^{-\frac{1}{2}D^2} \cdot I_i. 
  \label{eq:local}
\end{equation}
While this localized confinement might lead to a loss of information from more distant Gaussians, it is worth noting that all Gaussians are trainable and can compensate for any missing information from more distant Gaussians.

\noindent \textbf{Discretize and Align}
In reconstructing a 3D volume $V$, directly using the continuous mean $\mu$ for indexing and summing all Gaussian contributions presents a challenge: the gradient with respect to $\mu$ cannot propagate through this summation. This occurs because the discretized 3D volume $V$ relies on integer coordinates, while $\mu$ is continuous. Simply discretizing $\mu$ by rounding it to the nearest integer for indexing would render the reconstruction process non-differentiable, thereby hindering optimization.
To circumvent this, rather than directly discretizing $\mu$, we calculate each Gaussian's contribution at the discretized grid positions aligned with their \textit{floored} integer coordinates. More precisely, we select the floored integer of the box's center as the indexing point, denoting this integer-centered box as $\lfloor B \rfloor$.

For enhanced clarity, we define the residual between the continuous center $\mu$ and its floored integer counterpart as:
\begin{equation}
    \Delta \mu = \mu - \left \lfloor \mu \right \rfloor,
\end{equation}
where $\left \lfloor \mu \right \rfloor$ signifies the floor function. This $\Delta \mu$ quantifies the offset between the continuous and discretized positions, thereby enabling precise alignment between the continuous Gaussian center and the discretized grid.

Subsequently, we establish a relationship between the coordinates of the new box $\lfloor B \rfloor$ and the original box $B$ by accounting for this offset. The transformation from $B$ to $\lfloor B \rfloor$ is expressed as:
\begin{equation}
    \lfloor B\rfloor_{n,w_0,h_0,c_0,d}  = B_{w_0,h_0,c_0,d} - \Delta \mu_{n,1,1,1,d,}.
\end{equation}
Here, $\Delta \mu$ is implicitly broadcast across the dimensions of $B$ to enable element-wise subtraction. This adjustment guarantees that each Gaussian's contribution is accurately evaluated at its corresponding position on the discretized grid, relative to its floored mean. This `Discretize and Align' procedure thus enables seamless optimization by ensuring the unimpeded flow of gradients through the aggregation of Gaussian contributions, thereby preserving the differentiability of the reconstruction pipeline.

\vspace{-1mm}
\subsection{Fast Volume Reconstruction}
\vspace{-1mm}
\label{sec:recon}
\noindent \textbf{Parallel Computation}
To improve computational efficiency, we compute the squared Mahalanobis distance $D^2$ for each local region in parallel. For each voxel, $D^2$ is calculated using the discretized box $\lfloor B \rfloor$, leading to:
\begin{equation}
    D^2 = \lfloor B \rfloor^{\top}C^{-1}\lfloor B \rfloor.
\end{equation}
To represent this computation more clearly and compactly, we utilize the Einstein summation convention, which implicitly sums over repeated indices. The expression for $D^2$ thus becomes:
\begin{equation}
\label{eins}
D_{n,w_0,h_0,c_0}^2= \sum_{d}^{} \lfloor B \rfloor_{n,w_0,h_0,c_0,d} C^{-1}_{n,d,d} \lfloor B \rfloor_{n,w_0,h_0,c_0,d}. \nonumber
\end{equation}
Once $D^2$ is computed, the contributions of all Gaussians are summed to form the reconstructed volume.

\begin{figure*}[t]
  \centering
  \setlength{\belowcaptionskip}{-4mm}
  \includegraphics[width=.985\textwidth]{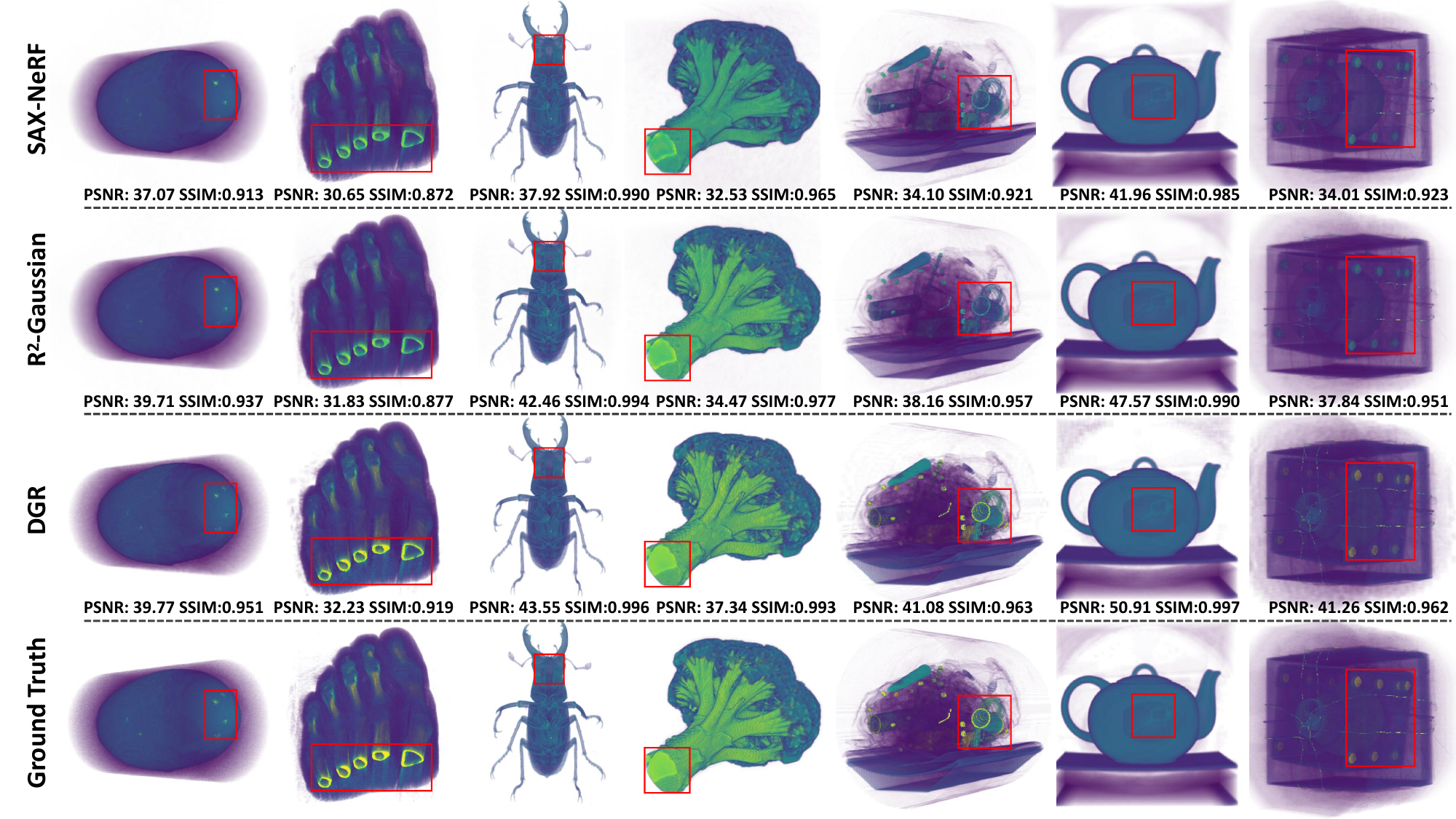}
  \caption{Qualitative Comparison of our DGR with previous state-of-the-art instance reconstruction methods SAX-NeRF and R$^2$-Gaussian, with major differences highlighted in \textcolor{red}{\textbf{Red}} boxes. \textit{Please zoom in for better visibility.}}
  \label{fig:visualization}
\end{figure*}

\noindent \textbf{Decomposition}
To further reduce computation in Equation~\ref{eins}, we decompose the computationally expensive Einstein summation into smaller, more manageable summations. The squared Mahalanobis distance $D^2$ can be explicitly decomposed into a combination of the following four smaller Einstein sums:
\begin{flalign}
& B^TC^{-1}B_{n,w_0,h_0,c_0} = \sum_{d}^{} B_{w_0,h_0,c_0,d}^T\ C^{-1}_{n,d,d} \ B_{w_0,h_0,c_0,d}, \nonumber \\
& B^TC^{-1}\Delta \mu_{n,w_0,h_0,c_0} =  \sum_{d}^{}  B_{w_0,h_0,c_0,d}^T\ C^{-1}_{n,d,d} \ \Delta \mu_{n,1,1,d}, \nonumber \\
&  \Delta \mu^TC^{-1}B_{n,w_0,h_0,c_0} =  \sum_{d}^{} \Delta \mu_{n,1,1,d}^T\ C^{-1}_{n,d,d} \ B_{w_0,h_0,c_0,d}, \nonumber \\
&  \Delta \mu^TC^{-1}\Delta \mu_{n,1,1,1} =  \sum_{d}^{} \Delta \mu_{n,1,1,d}^T\ C^{-1}_{n,d,d} \ \Delta \mu_{n,1,1,d}, \nonumber
\end{flalign}
Subsequently, we compute the squared Mahalanobis distance $D^2$ as:
\begin{align}
    D^2 = & (B^\top C^{-1}B) - (B^\top C^{-1}\Delta \mu) \nonumber \\
    & - (\Delta \mu^\top C^{-1}B) + (\Delta \mu^\top C^{-1}\Delta \mu). \label{eq:D2_decomposed}
\end{align}
The final contribution of the Gaussians within the local region is computed as:
\begin{equation}
    \Gamma = e^{-\frac{1}{2}D_{n,w_0,h_0,c_0}^2} \cdot I,
\end{equation}
where $\Gamma \in \mathbb{R}^{n\times w_0 \times h_0 \times c_0}$ represents the contributions of $n$ Gaussians within the local region. This decomposition allows us to calculate the contributions in a highly efficient manner, reducing time complexity without increasing space complexity. Specifically, $B_{w_0,h_0,c_0,d}$ is only $\frac{1}{n}$ the size of $\lfloor B \rfloor_{n,w_0,h_0,c_0,d}$, enabling far more efficient computations. The quantitative evidence supporting this efficiency gain is presented in Table~\ref{tab:space_time}.


Then we aggregate the contribution of each Gaussian into its corresponding voxels within the volume. Specifically, for each contribution $\Gamma_{i,x,y,z}$ from the $i$-th Gaussian to the voxel located at position $(x,y,z)$, the volume update proceeds as:
\begin{equation}
    V_{x,y,z} \leftarrow V_{x,y,z} + \Gamma_{i,x,y,z}.
\end{equation}
This aggregation is performed for all Gaussians, guaranteeing the precise accumulation of their contributions at the appropriate locations within the reconstructed volume. A highly parallelized implementation of this update process is available in our codebase for further reference.

Notably, our Fast Volume Reconstruction is a plug-and-play module that can be seamlessly integrated into existing methods leveraging 3D Gaussians. For instance, while X-Gaussian~\cite{x-gaussian} originally focuses on novel view synthesis without volume reconstruction, integrating our module into its pipeline prior to loss computation enables it to perform volume reconstruction, as demonstrated in Table~\ref{tab:real}.


\vspace{-1mm}
\subsection{Global Optimization}
\vspace{-1mm}
\label{sec:optimization}
Subsequent to reconstructing the 3D volume from the Gaussians, we adopt a joint optimization of the Gaussian parameters within the projection space. Diverging from conventional 3DGS and its variants, which typically optimize parameters per selected view~\cite{3DGS, x-gaussian} or local blocks~\cite{r2-gaussian}, our global optimization strategy achieves convergence in fewer than 1K iterations, representing a substantial improvement over standard approaches (e.g., requiring 30K iterations).

The initial phase of this optimization involves projecting the reconstructed 3D volume back into the projection domain. To accomplish this, we define a general geometry projection transformation $\mathcal{T}$:
\begin{equation}
    \hat P = \mathcal{T}(V),
\end{equation}
where $\hat P \in \mathbb{R}^{m \times n \times p}$ denotes the estimated projection.

This concise formulation is independent of specific CT acquisition geometries, demonstrating strong compatibility with various CT settings, including Fan-Beam or Cone-Beam CT, as well as Sparse-View or Limited-Angle CT.
In each iteration, we optimize the Gaussian parameters to minimize the discrepancy between the estimated projection $\hat P$ and the reference projection $P$. This optimization is guided by an objective function combining $\mathcal{L}_1$ loss, SSIM~\cite{ssim} loss ($\mathcal{L}_{SSIM}$), and total variation~\cite{tv} loss ($\mathcal{L}_{TV}$):
 \begin{equation}
        \mathcal{L}_{\text{total}} = \lambda_1 \mathcal{L}_{1}(\hat{P}, P) \!+\! \lambda_2 \mathcal{L}_{\text{SSIM}}(\hat{P}, P) \!+\! \lambda_3 \mathcal{L}_{\text{TV}}(V),
\end{equation}
where $\lambda_1=0.6$, $\lambda_2=0.2$, and $\lambda_3=1$ are the respective weights for the $\mathcal{L}_{1}$, $\mathcal{L}_{\text{SSIM}}$, and $\mathcal{L}_{\text{TV}}$. We discuss the definition and usage of these loss functions in Appendix \textcolor{iccvblue}{D.1}.

To further enhance the optimization process, we adopt the well-established adaptive density control for 3D Gaussians~\cite{3DGS}. This involves cloning under-reconstructed Gaussians, splitting over-reconstructed ones into smaller Gaussians, and pruning those with near-zero gradient magnitudes. As this process is standard and not our core contribution, we provide full details in Appendix~\textcolor{iccvblue}{F}.

\begin{table*}[t]
  \caption{Comparisons of 75/50/25-view Sparse-View CT on FIPS dataset. Best in \textbf{Bold.} }
  \centering
  \setlength{\tabcolsep}{2.1mm}
    \begin{tabular}{@{}lccccccccc@{}}
    \toprule
    \multirow{2}{*}{Methods} & \multicolumn{3}{c}{75-view} & \multicolumn{3}{c}{50-view} & \multicolumn{3}{c}{25-view} \\ \cmidrule(l){2-10} 
     & PSNR$\uparrow$ & SSIM$\uparrow$ & Time$\downarrow$ & PSNR$\uparrow$ & SSIM$\uparrow$ & Time$\downarrow$ & PSNR$\uparrow$ & SSIM$\uparrow$ & Time$\downarrow$ \\ \midrule
    \multicolumn{10}{c}{Real-World Dataset} \\ \midrule
    FDK~\cite{fdk} & 30.03 & 0.535 & - & 27.38 & 0.449 & - & 23.30 & 0.335 & - \\
    IntraTomo~\cite{intratomo} & 36.79 & 0.858 & 2h25m & 36.99 & 0.854 & 2h19m & \textbf{35.85} & 0.835 & 2h18m \\
    NAF~\cite{naf} & 38.58 & 0.848 & 51m28s & 36.44 & 0.818 & 51m31s & 32.92 & 0.772 & 51m24s \\
    SAX-NeRF~\cite{sax} & 34.93 & 0.854 & 13h21m & 34.89 & 0.840 & 13h23m & 33.49 & 0.793 & 13h25m \\
    X-Gaussian*~\cite{x-gaussian} & 38.27 & 0.894 & 10m21s & 37.80 & 0.881 & 10m5s & 35.12 & 0.859 & 9m55s \\
    R$^2$-Gaussian (iter=10k)~\cite{r2-gaussian} & 38.10 & 0.872 & 3m39s & 37.52 & 0.866 & 3m37s & 35.10 & 0.840 & 3m23s \\
    R$^2$-Gaussian (iter=30k)~\cite{r2-gaussian} & 39.40 & 0.875 & 14m16s & 38.24 & 0.864 & 13m52s & 34.83 & 0.833 & 12m56s \\ \hdashline
    DGR (iter=300) & 39.91 & 0.937 & \textbf{3m36s} & 38.66 & 0.929 & \textbf{3m24s} & 35.16 & \textbf{0.883} & \textbf{2m59s} \\
    DGR (iter=1000) & \textbf{41.28} & \textbf{0.952} & 13m14s & \textbf{39.27} &\textbf{ 0.940} & 11m48s & 34.58 & 0.870 & 8m54s \\ \midrule
    \multicolumn{10}{c}{Synthesis Dataset} \\ \midrule
    FDK~\cite{fdk} & 28.63 & 0.497 & - & 26.50 & 0.422 & - & 22.99 & 0.317 & - \\
    IntraTomo~\cite{intratomo} & 35.42 & 0.924 & 2h7m & 35.25 & 0.923 & 2h9m & 34.68 & 0.914 & 2h19m \\
    NAF~\cite{naf} & 37.84 & 0.945 & 30m43s & 36.65 & 0.932 & 32m4s & 33.91 & 0.893 & 31m1s \\
    SAX-NeRF~\cite{sax} & 38.07 & 0.950 & 13h5m & 36.86 & 0.938 & 13h5m & 34.33 & 0.905 & 13h3m \\
    X-Gaussian*~\cite{x-gaussian} & 38.22 & 0.955 & 9m58s & 37.50 & 0.946 & 9m50s & 35.18 & 0.924 & 9m51s \\
    R$^2$-Gaussian (iter=10k)~\cite{r2-gaussian} & 38.29 & 0.954 & 2m38s & 37.63 & 0.949 & 2m35s & 35.08 & 0.922 & 2m35s \\
    R$^2$-Gaussian (iter=30k)~\cite{r2-gaussian} & 38.88 & 0.959 & 8m21s & 37.98 & 0.952 & 8m14s & 35.19 & 0.923 & 8m28s \\ \hdashline
    DGR (iter=300) & 38.71 & 0.957 & \textbf{2m24s} & 38.05 & 0.954 & \textbf{2m17s} & 35.76 & 0.930 & \textbf{2m14s}\\
    DGR (iter=1000) & \textbf{39.63} & \textbf{0.964} & 7m27s & \textbf{38.74} & \textbf{0.960} & 7m13s & \textbf{36.55} & \textbf{0.947} & 6m53s \\ \bottomrule
    \end{tabular}
  \label{tab:real}
  \vspace{-2mm}
  \end{table*}

\section{Experiments}
\vspace{-1mm}
\subsection{Experiment Settings}
\vspace{-1mm}
\noindent \textbf{Datasets and Evaluation Setup}
We conducted experiments using three distinct datasets: FIPS~\cite{fips}, AAPM-Mayo LDCT~\cite{mayo}, and FUMPE~\cite{fumpe}. To ensure a fair and reproducible evaluation, we maintained consistent settings for all compared methods and grouped them by task, as detailed below:
For Cone-Beam Sparse-View CT, we evaluated on the FIPS~\cite{fips} dataset, following the setup used by R$^2$-Gaussian~\cite{r2-gaussian}. This dataset includes both real-world and synthetic data. The real-world evaluation utilized three cases with 721 real projections, while the synthetic portion involved 15 real CT volumes encompassing a range of subjects from organisms to artificial objects.
For Fan-Beam Sparse-View CT and Limited-Angle CT, we followed the evaluation protocols of SWORD~\cite{SWORD} and DiffusionMBIR~\cite{DiffusionMBIR}, using the widely adopted AAPM-Mayo LDCT dataset. This dataset comprises 5388 slices with a 1mm thickness, commonly used in CT reconstruction research. Additionally, we performed evaluations on the FUMPE~\cite{fumpe} dataset, which contains 8792 chest CT images. Following standard practice, we adopted Peak Signal-to-Noise Ratio (PSNR)~\cite{psnr} and Structural Similarity Index Measure (SSIM)~\cite{ssim} as our image quality metrics.

\noindent \textbf{Implementation Details}
We implemented DGR using PyTorch. Most experiments were trained on a single NVIDIA RTX 6000 GPU. However, for the 75/50/25-view Sparse-View CT experiments on the FIPS dataset, we utilized a single NVIDIA RTX 3090 GPU to ensure a fair comparison by maintaining the same training environment as R$^2$-Gaussian.
We initialized the number of Gaussians to 150K and set the local box size to $17\times 17\times 17$. The Adam optimizer~\cite{adam} was employed with an initial learning rate of $3 \times 10^{-4}$, which decayed to $3 \times 10^{-5}$ by the end of training. For Cone-Beam Sparse-View CT, the training process was limited to 1K iterations. For Fan-Beam Sparse-View CT, we did not set the iteration limit, and the training process is ceased when the validation loss converges in the projection domain.

\begin{table*}[t]
\setlength{\abovecaptionskip}{2mm}
  \caption{Comparisons of 180/120/90/60-view Sparse-View CT on AAPM-Mayo LDCT dataset. Best in \textbf{Bold.}}
  \centering
  \setlength{\tabcolsep}{2.5mm}
  \begin{tabular}{lccccccccc}
  \toprule
  \multicolumn{1}{c}{\multirow{2}{*}{Method}} & \multirow{2}{*}{Extra Data} & \multicolumn{2}{c}{180-view} & \multicolumn{2}{c}{120-view} & \multicolumn{2}{c}{90-view} & \multicolumn{2}{c}{60-view} \\ 
  \cline{3-10} 
   &  & PSNR$\uparrow$& SSIM$\uparrow$ & PSNR$\uparrow$ & SSIM$\uparrow$ & PSNR$\uparrow$ & SSIM$\uparrow$ & PSNR$\uparrow$ & SSIM$\uparrow$ \\ 
  \hline
  FBP~\cite{fbp1967} & 0 & 31.69 & 0.882 & 28.30 & 0.787 & 26.20 & 0.701 & 23.18 & 0.595 \\ 
  FBPConvNet~\cite{FBPConvNet} & 4839 & 42.23 & 0.988 & 39.45 & 0.983 & 37.11 & 0.976 & 35.63 & 0.966 \\
  U-Net~\cite{UNetct} & 4839 & 38.37 & 0.985 & 35.58 & 0.977 & 30.09 & 0.947 & 28.83 & 0.937 \\
  PLANet~\cite{PLA} & 4176 & 42.76 & 0.965 & 41.67 & 0.962 & 40.99 & 0.957 & 38.97 & 0.941 \\
  GMSD~\cite{GMSD} & 4839 & 41.44 & 0.988 & 39.41 & 0.981 & 37.25 & 0.974 & 34.31 & 0.958 \\
  SWORD~\cite{SWORD} & 4839 & 45.08 & 0.994 & 42.49 & 0.990 & 41.27 & 0.986 & 38.49 & 0.978 \\
  DGR & 0 & \textbf{46.13} & \textbf{0.997} & \textbf{44.64} & \textbf{0.994} & \textbf{43.23} & \textbf{0.992} & \textbf{40.25} & \textbf{0.985} \\ 
  \bottomrule
  \end{tabular}
  \label{tab:svct}
  \vspace{-3mm}
  \end{table*}

  \begin{table}[t]
  \setlength{\abovecaptionskip}{2mm}
  \caption{ Fan-Beam Limited-Angle CT 90$^\circ$ reconstruction.}
  \scriptsize
  \centering
  \setlength{\tabcolsep}{0.7mm}
  \begin{tabular}{lcccccccc}
  \hline
  \multicolumn{1}{c}{\multirow{2}{*}{Method}} & \multicolumn{1}{c}{\multirow{2}{*}{Data}} & \multicolumn{2}{c}{Axial} & \multicolumn{2}{c}{Coronal} & \multicolumn{2}{c}{Sagittal} \\ 
  \cline{3-8}
  & & PSNR $\uparrow$ & SSIM $\uparrow$ & PSNR $\uparrow$ & SSIM $\uparrow$ & PSNR $\uparrow$ & SSIM $\uparrow$ \\ \hline
  FBP~\cite{fbp1967} & 0 & 14.91 & 0.397 & 17.07 & 0.411 & 15.46 & 0.403 \\ 
  FBPConvNet~\cite{FBPConvNet} & 3142 & 26.76 & 0.879 & 25.77 & 0.874 & 22.92 & 0.841 \\
  ADMM-TV~\cite{ADMM-TV} & 3142 & 23.19 & 0.793 & 22.96 & 0.758 & 19.95 & 0.782 \\
  MCG~\cite{MCG} & 3839 & 26.01 & 0.838 & 24.55 & 0.823 & 21.59 & 0.706 \\
  Lahiri \textit{et al.}~\cite{lahiri2023sparse} & 3142 & 28.08 & 0.931 & 26.02 & 0.856 & 23.24 & 0.812 \\
  DiffusionMBIR~\cite{DiffusionMBIR} & 3142 & 34.92 & 0.956 & 32.48 & 0.947 & 28.82 & 0.832 \\
  DGR & 0 & \textbf{38.22} & \textbf{0.970} & \textbf{39.32} & \textbf{0.969} & \textbf{38.35} & \textbf{0.970} \\
  \bottomrule
  \end{tabular}
  \label{tab:lact}
  \vspace{-5mm}
  \end{table}

\vspace{-1mm}
\subsection{Experimental Results and Analysis}
\vspace{-1mm}
\label{exp}

\subsubsection{Cone-Beam Sparse-View CT Reconstruction}
In Table \ref{tab:real}, we present a comparison of DGR with advanced instance reconstruction methods on the FIPS dataset. Baselines include NeRF-based methods (IntraTomo~\cite{intratomo}, NAF~\cite{naf}, SAX-NeRF~\cite{sax}) and 3DGS-based methods (X-Gaussian~\cite{x-gaussian}, R$^2$-Gaussian~\cite{r2-gaussian}), with X-Gaussian specifically implemented using our Fast Volume Reconstruction module. DGR demonstrates superior image quality and computational efficiency across all view counts, significantly outperforming previous instance reconstruction methods. Even with just 300 iterations, DGR surpasses most advanced methods while completing reconstruction in approximately 3 minutes.

The faster reconstruction time achieved by DGR is attributed to the discretized representation, which is inherently well-aligned with the reconstruction task and avoids redundant computation. Furthermore, DGR's lightweight framework enables the joint optimization of all Gaussians, significantly reducing computation time without compromising reconstruction quality, as illustrated in Figure~\ref{fig:visualization}.

\vspace{-1mm}
\subsubsection{Fan-Beam Sparse-View CT Reconstruction}
In Table \ref{tab:svct}, we compare DGR with leading DLR methods on the AAPM-Mayo LDCT dataset~\cite{mayo}. The term `Data' in these tables refers to the number of images used to train the DLR methods, whereas our DGR optimizes without relying on additional training data.
The baseline methods include traditional CNN-based approaches such as FBPConvNet~\cite{FBPConvNet}, U-Net~\cite{UNetct}, and PLANet~\cite{PLA}, alongside advanced diffusion-based methods like GMSD~\cite{GMSD}, and SWORD~\cite{SWORD}. To ensure a fair comparison for 180/120/90/60-view Sparse-View CT, we replicated the settings used in SWORD~\cite{SWORD}.
Notably, DGR surpasses advanced DLR methods that incorporate prior knowledge in Fan-Beam Sparse-View CT reconstruction, despite requiring zero training data. This validates DGR's ability to address diverse CT reconstruction tasks without architectural modifications while achieving superior performance.

\vspace{-1mm}
\subsubsection{Fan-Beam Limited-Angle CT Reconstruction}
In Table \ref{tab:lact}, we report the performance of Limited-Angle CT reconstruction experiments conducted on the AAPM-Mayo LDCT dataset~\cite{mayo}, following the methodology of DiffusionMBIR~\cite{DiffusionMBIR}. The baseline methods include traditional CNN-based techniques such as FBPConvNet~\cite{FBPConvNet} and Lahiri \textit{et al.}~\cite{lahiri2023sparse}, alongside advanced diffusion-based approaches like MCG~\cite{MCG} and DiffusionMBIR~\cite{DiffusionMBIR}. For these evaluations, the angular data acquisition was restricted to a 90$^\circ$ range within a total of 180$^\circ$. Without any prior knowledge, DGR still demonstrates a substantial improvement over these advanced DLR methods.

\vspace{-1mm}
\subsubsection{Additional Evaluations and Visualizations}
Our DGR is further evaluated on the FUMPE dataset~\cite{fumpe} for Fan-Beam 180/120/90/60-view chest CT reconstruction, with the quantitative results detailed in Appendix Table \textcolor{iccvblue}{9}. Furthermore, iterative reconstruction visualizations of DGR are showcased in Appendix Figures \textcolor{iccvblue}{4}, \textcolor{iccvblue}{5}, and \textcolor{iccvblue}{6}.

\subsection{Ablation Study}
In Table \ref{tab:box_size}, we evaluate the impact of local box size on 60-view Sparse-View CT reconstruction using the AAPM-Mayo LDCT dataset. Each configuration was assessed over 1K iterations to ensure stable measurements of training time and video memory usage. We observe that performance improves with increasing box size; however, this comes at the cost of increased time consumption and video memory. Consequently, we chose $17\times 17\times 17$ as the default box size for our experiments, as it achieves an optimal balance between performance and efficiency.
\begin{table}[t]
  \scriptsize
  \centering
  \setlength{\abovecaptionskip}{2mm}
  \caption{Effect of Box Size on 60-view Sparse-View CT}
  \setlength{\tabcolsep}{0.7mm}
  \begin{tabular}{@{}lcccc@{}}
  \toprule
  Box-Size & $13\times 13\times 13$ & $15\times 15\times 15$ & $17\times 17\times 17$ & $19\times 19\times 19$\\ \midrule
  Time (minutes) & 7.92 & 11.07 & 16.58 & 26.99 \\
  V-Ram (GiB) & 10.70 & 13.29 & 16.87 & 21.32  \\
  PSNR / SSIM & 35.99 / 0.960 & 38.90 / 0.973 & 40.25 / 0.985 & 40.98 / 0.987  \\ \bottomrule
  \end{tabular}
  \label{tab:box_size}
  \vspace{-5mm}
  \end{table}

\noindent \textbf{Effect of Fast Volume Reconstruction} 
In Table \ref{tab:space_time}, we compare the video memory and time consumption of our Fast Volume Reconstruction (denoted as FVR below) against the Direct reconstruction (Equation \ref{eq:gaussian}). We observe that Fast Volume Reconstruction achieves a significant reduction in video memory consumption. The decomposition further reduces the reconstruction time from 1.05s to 0.09s, thereby enabling the highly efficient reconstruction capabilities.

   \begin{table}[h]
    \centering
    \setlength{\abovecaptionskip}{2mm}
    \setlength{\belowcaptionskip}{-2mm}
    \caption{ Video-Memory and Time Per Iteration}
    \footnotesize
    \setlength{\tabcolsep}{3.2mm}
    \begin{tabular}{lcc}
    \toprule
    \multicolumn{1}{c}{Method} & VRAM  (GiB) & Time (s) \\ 
    \midrule
    Direct Reconstruction (estimated) &  16662.50 & / \\
    FVR w/o Decomposition & 16.87 & 1.05 \\
    FVR w/ Decomposition & 16.87 & 0.09 \\ 
    \bottomrule
    \end{tabular}
    \label{tab:space_time}
    \vspace{-2mm}
    \end{table}
\noindent \textbf{Analysis of Loss Functions}
In Appendix Tables \textcolor{iccvblue}{6}, \textcolor{iccvblue}{7}, and \textcolor{iccvblue}{8}, we discuss the usage of various loss functions. Results indicate that a combined use of $\mathcal{L}_1$ loss, SSIM loss ($\mathcal{L}_{SSIM}$), and total variation loss ($\mathcal{L}_{TV}$) yields the best performance.

\noindent \textbf{Analysis of Isotropic Gaussians and Voxelizer}
As detailed in Appendix Sections \textcolor{iccvblue}{D.2} and \textcolor{iccvblue}{D.3}, our ablation studies demonstrate that DGR surpasses alternative designs in both image quality and reconstruction speed.

\vspace{-1mm}
\section{Conclusion}
\vspace{-1mm}
We present DGR, a novel framework that rethinks CT reconstruction through representation, reconstruction, and optimization. Evaluated on multiple datasets against both Deep Learning Reconstruction and instance reconstruction baselines, DGR achieves superior image quality and speed, demonstrating significant potential for clinical application.

\clearpage
\section*{Acknowledgments}
This research is supported by the Fundamental Research Funds for the Central Universities (project number YG2024ZD06), NSFC (No. 62176155), and Shanghai Municipal Science and Technology Major Project (2021SHZDZX0102).

{
    \small
    \bibliographystyle{ieeenat_fullname}
    \bibliography{egbib}
}
\maketitle

\begin{appendix}
	
	\begin{center}
		{\Large \bf Appendix}
	\end{center}
	
\end{appendix}

\appendix

\section{Overview}
Thank you for reading the Appendix of our research. This appendix is organized as follows:
\begin{itemize}
    \item Section \ref{sec:detail} provides a detailed explanation of the experimental setup.
    \item Section \ref{sec:metrics} describes the evaluation metrics used in our study.
    \item Section \ref{sec:loss} discusses the application of different loss functions, types of Gaussians, and voxelizers.
    \item Section \ref{sec:fast} elaborates on the Fast Volume Reconstruction method and analyzes its computational complexity.
    \item Section~\ref{densify} details the implementation of adaptive density control in this work.
    \item Section \ref{sec:vis} presents additional visualizations of DGR reconstruction.
\end{itemize}

\section{Experiment Details}
\label{sec:detail}

\subsection{Code and Reproducibility}
Our code is publicly available at \url{https://github.com/wskingdom/DGR} and is meticulously organized for readability. Experiments are categorized into three groups, each with corresponding code: (1) Cone-Beam Sparse-View CT, (2) Fan-Beam Sparse-View CT, and (3) Fan-Beam Limited-Angle CT. To reproduce our results, please refer to the \textbf{Readme.md} file in our codebase.

\subsection{Organization of Experiments}
To accommodate diverse experimental settings in related research (e.g., varying projection numbers and scanning geometry), our experiments are divided into three groups:

\noindent \textbf{Cone-Beam Sparse-View CT:} Experiments with 75, 50, and 25 views were conducted on the FIPS dataset~\cite{fips}, aligning with the settings of R$^2$-Gaussian~\cite{r2-gaussian}.

\noindent \textbf{Fan-Beam Sparse-View CT:} Experiments with 180, 120, 90, and 60 views were performed on the AAPM-Mayo LDCT dataset~\cite{mayo} and the FUMPE dataset~\cite{fumpe}, following the methodology of the advanced Deep Learning Reconstruction (DLR) method, SWORD~\cite{SWORD}.

\noindent \textbf{Fan-Beam Limited-Angle CT:} Experiments on 90$^\circ$ Limited-Angle CT were conducted on the AAPM-Mayo LDCT dataset~\cite{mayo}, consistent with  DiffusionMBIR~\cite{DiffusionMBIR}.

\section{Evaluation Metrics}
\label{sec:metrics}
\subsection{Peak Signal-to-Noise Ratio (PSNR)}
PSNR~\cite{psnr} is a widely used metric to evaluate the quality of the reconstructed images. It is defined as:
\begin{equation}
    \text{PSNR}(x, y) = 10 \cdot \log_{10} \left( \frac{{\text{MAX}^2}}{{\text{MSE}(x, y)}} \right),
\end{equation}
where $\text{MAX}$ is the maximum possible pixel value of the image and $\text{MSE}(x, y)$ is the mean squared error between the original and reconstructed images.

\subsection{Structural Similarity Index (SSIM)}
SSIM~\cite{ssim} is a metric that measures the similarity between two images. It is defined as:
\begin{equation}
    \text{SSIM}(x, y) = \frac{(2\mu_x\mu_y + C_1)(2\sigma_{xy} + C_2)}{(\mu_x^2 + \mu_y^2 + C_1)(\sigma_x^2 + \sigma_y^2 + C_2)},
 \end{equation} 
where $\mu_x$ and $\mu_y$ are the mean values of the images $x$ and $y$, $\sigma_x^2$ and $\sigma_y^2$ are the variances of the images, $\sigma_{xy}$ is the covariance of the images, and $C_1$ and $C_2$ are constants to stabilize the division when the denominator is small.

  \begin{table}[t]
  \centering
  \caption{Usage of Different Loss Functions}
\label{loss-cbct}
  \begin{tabular}{@{}ccccc@{}}
  \toprule
  \multicolumn{5}{c}{Cone-Beam Sparse-View CT (50 view, 300 iter)} \\ \midrule
  $\mathcal{L}_{1}$ Loss & SSIM Loss & TV Loss & PSNR & SSIM \\
  \checkmark &  &  & 38.66 & 0.929 \\
  \checkmark & \checkmark &  & 38.94 & 0.933 \\
  \checkmark & \checkmark & \checkmark & \textbf{39.65} & \textbf{0.939} \\ \bottomrule
  \end{tabular}
  \end{table}

\begin{table}[t]
  \centering
  \caption{Usage of Different Loss Functions}
\label{loss-svct}
  \begin{tabular}{@{}ccccc@{}}
  \toprule
  \multicolumn{5}{c}{Fan-Beam Sparse-View CT (60 view, 300 iter)} \\ \midrule
  $\mathcal{L}_{1}$ Loss & SSIM Loss & TV Loss & PSNR & SSIM \\
  \checkmark &  &  & 38.82 & 0.931 \\
  \checkmark & \checkmark &  & 38.88 & 0.933 \\
  \checkmark & \checkmark & \checkmark & \textbf{39.15} & \textbf{0.936} \\ \bottomrule
  \end{tabular}
  \end{table}

\begin{table}[t]
  \centering
  \caption{Usage of Different Loss Functions}
\label{loss-lact}
  \begin{tabular}{@{}ccccc@{}}
  \toprule
  \multicolumn{5}{c}{Fan-Beam Limited-Angle CT (90$^\circ$, 300 iter)} \\ \midrule
  $\mathcal{L}_{1}$ Loss & SSIM Loss & TV Loss & PSNR & SSIM \\
  \checkmark &  &  & 37.69 & 0.932 \\
  \checkmark & \checkmark &  & 37.81 & 0.934 \\
  \checkmark & \checkmark & \checkmark & \textbf{38.02} & \textbf{0.936} \\ \bottomrule
  \end{tabular}
  \vspace{-3mm}
  \end{table}

\begin{figure*}[t]
    \centering
    \includegraphics[width=\textwidth]{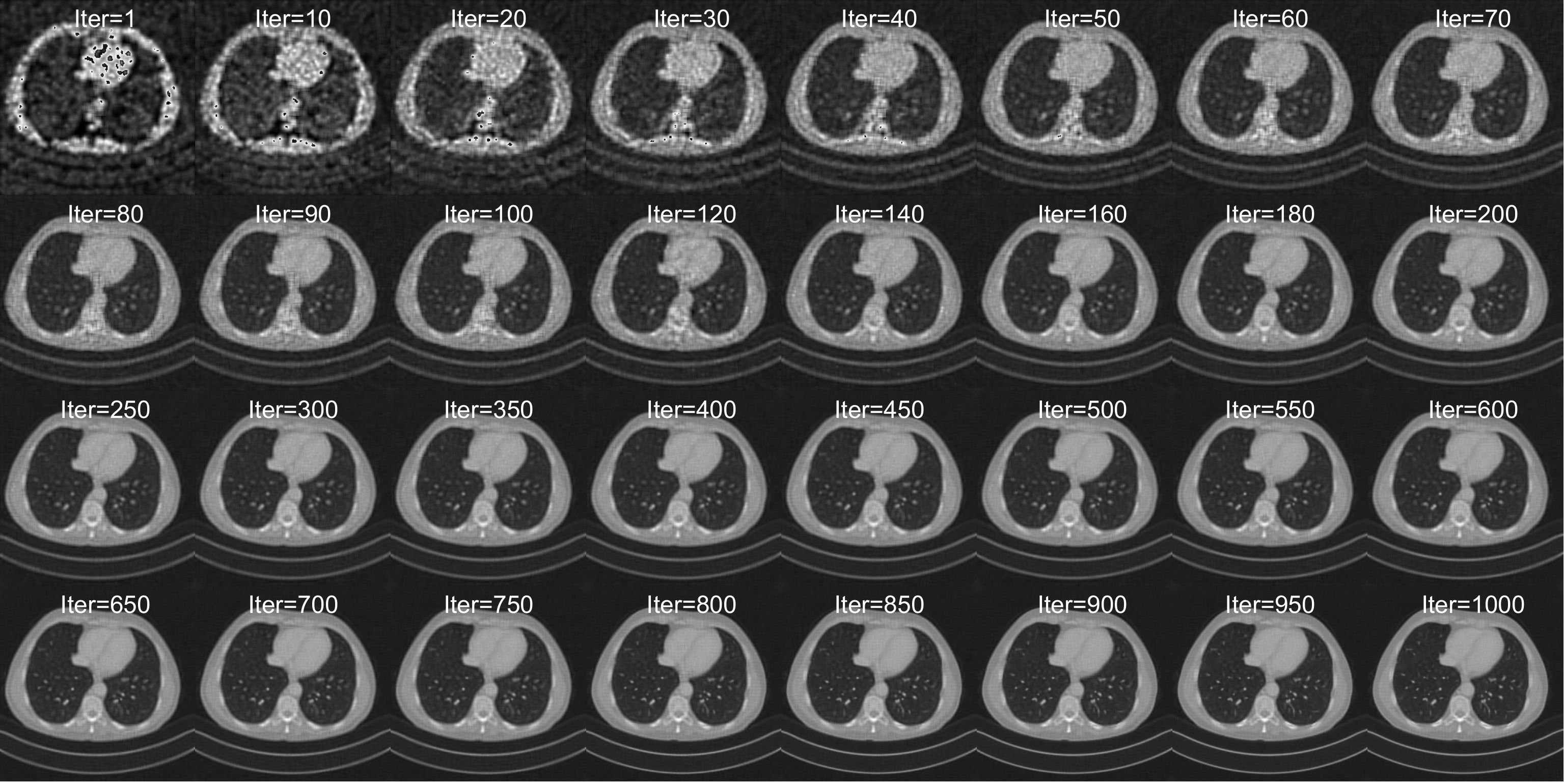}
    \caption{This figure illustrates the reconstruction process by iterations of Fan-Beam 60-view CT. The positions of Gaussians are initialized using Filtered Back Projection, as described in the experimental settings.}
    \label{fig:svct}
  \end{figure*}

\subsection{Evaluation Details}

For the Cone-Beam 75/50/25-view Sparse-View CT experiments, we assess performance on the FIPS dataset~\cite{fips} and maintain its original data split for consistency.

In the Fan-Beam 180/120/90/60-view Sparse-View CT experiments, we evaluate the performance of the reconstruction methods over the entire 3D volume, as established in SWORD~\cite{SWORD}. In this context, $x$ and $y$ in the PSNR and SSIM metrics refer to the reconstructed 3D volume and the ground truth 3D volume, respectively.

For the 90$^\circ$ Limited-Angle CT experiments, we evaluate performance across axial, coronal, and sagittal slices following DiffusionMBIR~\cite{DiffusionMBIR}. Here, $x$ and $y$ in the PSNR and SSIM metrics denote the reconstructed 2D slice and the ground truth 2D slice, respectively. The mean PSNR and SSIM values are then computed across all slices.

\begin{table*}[t]
  \caption{Comparisons of 180/120/90/60-view Sparse-View CT on FUMPE dataset. Best in \textbf{Bold.}}
  \centering
  \renewcommand\arraystretch{1}
  \setlength{\tabcolsep}{1.5mm}
  \begin{tabular}{lcccccccccc}
  \hline
  \multicolumn{1}{c}{\multirow{2}{*}{Method}} & \multicolumn{1}{c}{\multirow{2}{*}{Extra Data}} & \multicolumn{2}{c}{180-view} & \multicolumn{2}{c}{120-view} & \multicolumn{2}{c}{90-view} & \multicolumn{2}{c}{60-view}\\ 
  \cline{3-10}
  & & PSNR$\uparrow$& SSIM$\uparrow$& PSNR$\uparrow$& SSIM$\uparrow$& PSNR$\uparrow$& SSIM$\uparrow$& PSNR$\uparrow$& SSIM$\uparrow$\\ \hline
  FBP~\cite{fbp1967} & 0 & 33.74 & 0.859 & 30.65 & 0.803 & 28.52 & 0.757 & 25.59& 0.592\\
  MCG~\cite{MCG} & 8341 & 36.93 & 0.898 & 37.05 & 0.899 & 37.03 & 0.899 & 35.46 & 0.900 \\
  DiffusionMBIR~\cite{DiffusionMBIR} & 8341 & 36.78 & 0.914 & 36.50 & 0.907 & 36.79 & 0.918  & 34.86 & 0.918 \\
  SWORD~\cite{SWORD} & 8341 & 41.50 & 0.967 & 39.69 & 0.959 & 37.77 & 0.948 & 31.64 & 0.910\\
  DGR & 0 & \textbf{41.78} & \textbf{0.978} & \textbf{40.13} & \textbf{0.969} & \textbf{38.13} & \textbf{0.961} & \textbf{36.48} &  \textbf{0.923} \\ \bottomrule
  \end{tabular}
  \label{fumpe}
 \vspace{-3mm}
  \end{table*}

\section{Discussions}
\label{sec:loss}
\subsection{Ablation on Loss Functions}
In the main text, we use different loss functions together to enhance reconstruction quality. Specifically, the SSIM loss is computed in the projection domain to preserve structural information, while the TV loss is computed in the volume domain to promote sparsity in the reconstructed volume.
\begin{equation}
    \mathcal{L}_{\text{total}} = \lambda_1 \mathcal{L}_{1}(\hat{P}, P) + \lambda_2 \mathcal{L}_{\text{SSIM}}(\hat{P}, P), \nonumber
\end{equation}
where $\lambda_1=0.8$ and $\lambda_2=0.2$ are the weights of the $\mathcal{L}_{1}$ Loss and SSIM loss, respectively.

A combined use of $\mathcal{L}_{1}$ Loss, SSIM loss, and TV loss is as follows:
 \begin{equation}
        \mathcal{L}_{\text{total}} = \lambda_1 \mathcal{L}_{1}(\hat{P}, P) \!+\! \lambda_2 \mathcal{L}_{\text{SSIM}}(\hat{P}, P) \!+\! \lambda_3 \mathcal{L}_{\text{TV}}(V), \nonumber
\end{equation}
where $\lambda_1=0.6$, $\lambda_2=0.2$, and $\lambda_3=1$ are the weights of the $\mathcal{L}_{1}$ Loss, SSIM loss, and TV loss, respectively.

The results for these loss combinations are presented in Table~\ref{loss-cbct}, Table~\ref{loss-svct}, and Table~\ref{loss-lact}, corresponding to Cone-Beam Sparse-View CT, Fan-Beam Sparse-View CT, and Fan-Beam Limited-Angle CT experiments on the real-world FIPS dataset~\cite{fips}. The results demonstrate that the combined use of $\mathcal{L}_{1}$ Loss, SSIM loss, and TV loss achieves the best performance in terms of PSNR and SSIM metrics.

\subsection{Isotropic Gaussians vs Anisotropic Gaussians}
The use of isotropic Gaussians is justified by the generally isotropic nature of tissue attenuation properties in CT, especially for soft tissues, which exhibit minimal directional dependence. As noted in Principles of Computerized Tomographic Imaging~\cite{kak}, small elements of a distributed source can be treated as isotropic, reflecting tissue behavior. 

To further validate this point, we conduct ablation studies on 50-view Sparse-View CT of Table 1 (Synthetic dataset), repeating experiments 10 times and reporting Mean $\pm$ SD in Table~\ref{anisotropic} to evaluate the result of substituting our isotropic Gaussians with  anisotropic Gaussians. Results also validate that using isotropic Gaussians can not only achieve better image quality and stability but also reduce training time.

  \begin{table}[ht]
  \centering
  \footnotesize
  \caption{Replace isotropic Gaussians by Anisotropic Gaussians}
   \label{anisotropic}
\begin{tabular}{@{}lccc@{}}
\toprule
 & PSNR$\uparrow$ & SSIM$\uparrow$ & Time$\downarrow$ \\ \midrule
DGR & \textbf{38.74$\pm$ 0.07} & \textbf{0.960$\pm$ 0.01} & \textbf{7m13s} \\
w/ anisotropic Gaussian & 38.71$\pm$ 0.17 & 0.960$\pm$ 0.02 & 8m05s \\ \bottomrule
\end{tabular}
\end{table}

\subsection{Comparison with Other Voxelizer}
We also conduct ablation studies on 50-view Sparse-View CT of Table 1 (Synthetic dataset) and repeat experiments 10 times and reporting Mean $\pm$ SD in Table~\ref{voxelizer} to evaluate the result of substituting our Fast Volume Reconstruction with the R$^2$-Gaussian voxelizer. Results show that our reconstruction module is faster than R$^2$-Gaussian voxelizer. Besides, our reconstruction module boosts the Gaussian optimization jointly, which achieves better performance.

  \begin{table}[ht]
  \centering
  \footnotesize
  \caption{Replace our voxelizer by R$^2$-Gaussian Voxelizer}
   \label{voxelizer}
\begin{tabular}{@{}lccc@{}}
\toprule
 & PSNR$\uparrow$ & SSIM$\uparrow$ & Time$\downarrow$ \\ \midrule
DGR & \textbf{38.74$\pm$ 0.07} & \textbf{0.960$\pm$ 0.01} & \textbf{7m13s} \\
w/ R$^2$-Gaussian Voxelizer & 38.02$\pm$ 0.22 & 0.956$\pm$ 0.02 & 8m37s \\ \bottomrule
\end{tabular}
\vspace{-3mm}
\end{table}

\section{Fast Volume Reconstruction}
\label{sec:fast}
\subsection{Highly Parallelized Implementation}
We show the detailed implementation of our Fast Volume Reconstruction In Algorithm \textcolor{iccvblue}{1} and our code. The algorithm takes the mean $\mu$, covariance $C$, and intensity $I$ of the Gaussians as input and reconstructs the 3D volume in a fast and efficient manner. Notably, this process is implemented in a highly parallelized manner to accelerate the computation without sacrificing the reconstruction quality. More details can be found in our implementation code.

\subsection{Analysis of Complexity}
    \begin{table}[ht]
      \centering
    \footnotesize
    \setlength{\tabcolsep}{3mm}
    \begin{tabular}{lcc}
    \toprule
    \multicolumn{1}{c}{Method} & VRAM  (GiB) & Time (s) \\ 
    \midrule
    Direct Reconstruction (estimated) &  16662.50 & / \\
    FVR w/o Decomposition & 16.87 & 1.05 \\
    FVR w/ Decomposition & 16.87 & 0.09 \\ 
    \bottomrule
    \end{tabular}
    \end{table}

\begin{figure*}[t]
    \centering
    \includegraphics[width=\textwidth]{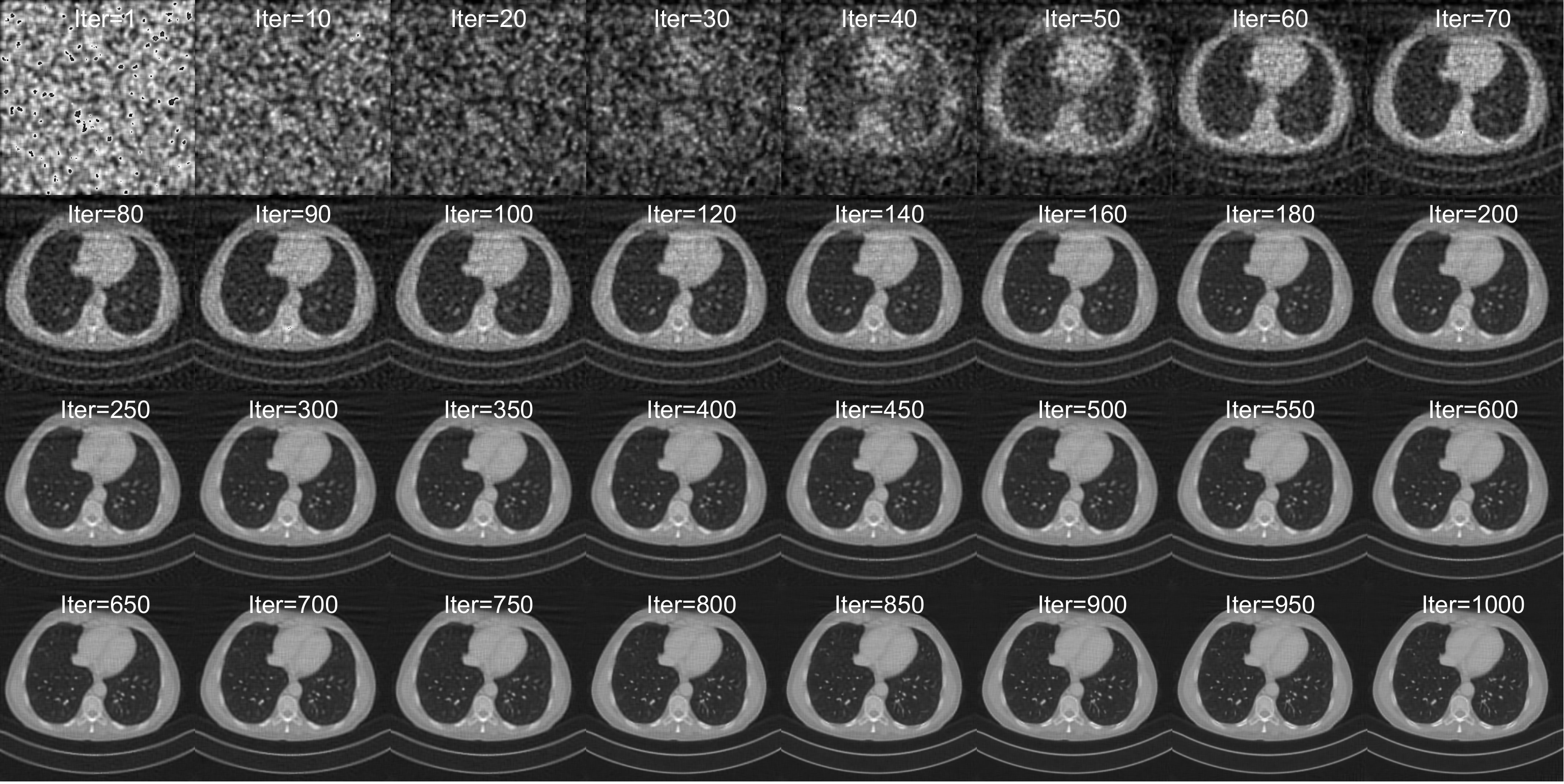}
    \caption{This figure illustrates the reconstruction process by iterations of Fan-Beam 90$^\circ$ Limited-Angle CT. The positions of Gaussians are randomly initialized for visualization comparison.}
    \label{fig:lact}
  \end{figure*}

\begin{figure*}[t]
    \centering
    \includegraphics[width=\textwidth]{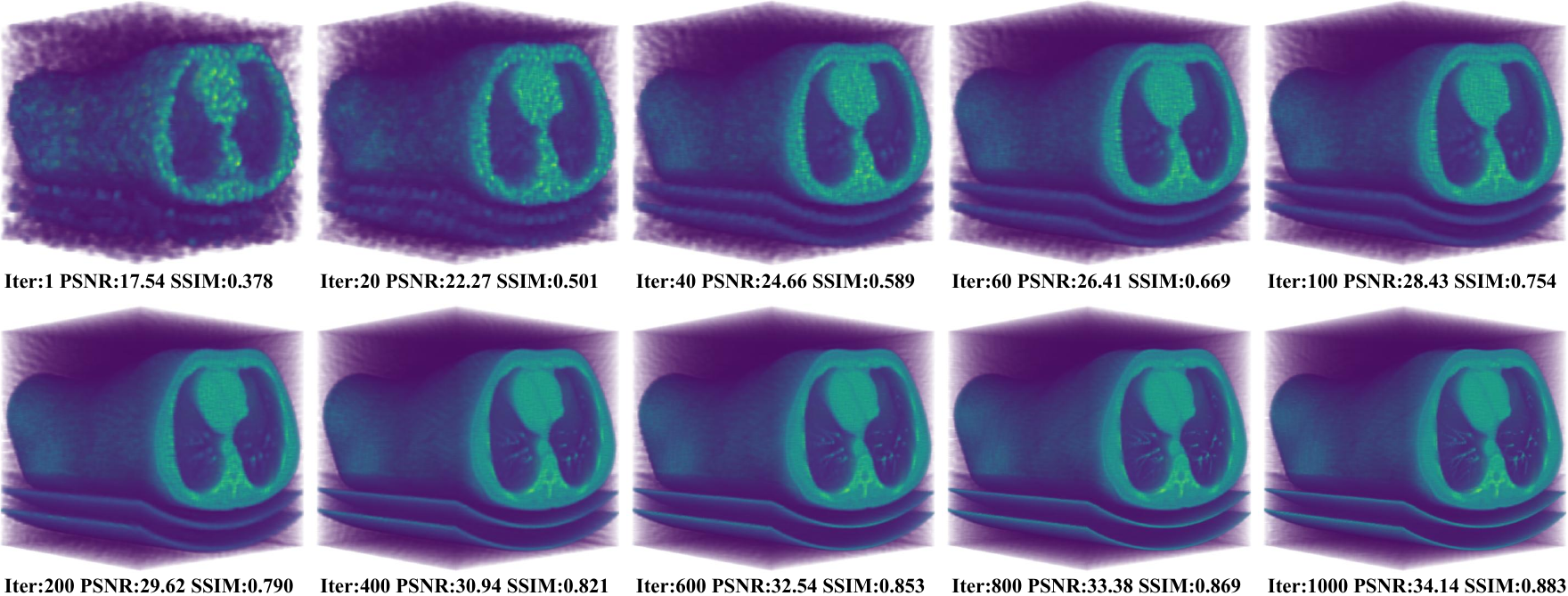}
    \caption{This figure illustrates the reconstruction process by iterations of Fan-Beam 60-view Sparse-View CT by 3D volume visualization.}
    \label{fig:iter}
  \end{figure*}
 
As mentioned in the ablation study, the proposed fast volume reconstruction saves both time and space consumption. This difference in complexity primarily stems from the computation of the Mahalanobis distance, $D^2$.

\noindent \textbf{Direct Reconstruction}
In the direct reconstruction approach, the Mahalanobis distance is calculated as $D^2 = (P-\mu)^\top\Sigma^{-1}(P-\mu)$. Here, $P \in \mathbb{R}^{w \times h \times c \times d}$ represents the position of a voxel in the volume, where $(d=3)$ indicates the spatial dimension. $\mu \in \mathbb{R}^{n \times d}$ is the set of means for $n$ Gaussians, and $\Sigma \in \mathbb{R}^{n \times d \times d}$ is the covariance matrix for these Gaussians. The computational complexity for direct reconstruction is $O(n \cdot w \cdot h \cdot c \cdot d^2)$. Evidently, this method is not feasible for a large number of Gaussians due to its prohibitive memory consumption.

\noindent \textbf{Fast w/o Decomposition}
In this condition, the Mahalanobis distance is calculated as $D^2 = \sum_{d}^{} B''_{n,w_0,h_0,c_0,d}\ C^{-1}_{n,d,d} \ B''_{n,w_0,h_0,c_0,d}$. Here, $w_0$, $h_0$, and $c_0$ denote the dimensions of the local region. The computational complexity for this approach is $O(n \cdot w_0 \cdot h_0 \cdot c_0 \cdot d^2)$. Compared to direct reconstruction, the local region size ($w_0 \cdot h_0 \cdot c_0$) is significantly smaller than the full volume size ($w \cdot h \cdot c$), leading to a substantial reduction in memory consumption.

\noindent \textbf{Fast Volume Reconstruction with Decomposition}
The decomposed form of Fast Volume Reconstruction yields the Mahalanobis distance as:
{\scriptsize
\[
D^2 = \sum_d \left( B'_{w_0,h_0,c_0,d} - \Delta \mu_{n,1,1,d} \right) C^{-1}_{n,d,d} \left( B'_{w_0,h_0,c_0,d} - \Delta \mu_{n,1,1,d} \right)
\]
}
The space complexity remains equivalent to that of fast volume reconstruction without decomposition. However, the time complexity is significantly reduced to $O(w_0 \cdot h_0 \cdot c_0 \cdot d)$, a benefit derived from the decomposition of the large matrix multiplication.


\begin{algorithm*}
  \caption{Fast Volume Reconstruction}
  \begin{algorithmic}
  \Require $\mu \in \mathbb{R}^{n \times d}$: mean of Gaussians ($d=3$ for 3D space)
  \Require $C \in \mathbb{R}^{n \times d \times d}$: covariance of Gaussians
  \Require $I \in \mathbb{R}^{n \times 1}$: intensity of Gaussians
  \State $V \gets \mathbf{O}_{w,h,c}$ \Comment Initialize target volume with zeros
  \State $B'_{w_0, h_0, c_0, d}\gets \texttt{meshgrid}(\{-\frac{w_0-1}{2}, \ldots, \frac{w_0-1}{2}\}, \{-\frac{h_0-1}{2}, \ldots, \frac{h_0-1}{2}\}, \{-\frac{c_0-1}{2}, \ldots, \frac{c_0-1}{2}\})$ \Comment Initialize the shift of local regions
  \State $\Delta \mu \gets \mu - \left \lfloor \mu \right \rfloor$ \Comment Align $\mu$ to discrete voxel grid
  \State $B'^TC^{-1}B'_{n,w_0,h_0,c_0} \gets \sum_{d}^{} B'_{w_0,h_0,c_0,d}\ C^{-1}_{n,d,d} \ B'_{w_0,h_0,c_0,d}$ \Comment Decompose large matrix multiplication
  \State $B'^TC^{-1}\Delta \mu_{n,w_0,h_0,c_0} \gets  \sum_{d}^{}  B'_{w_0,h_0,c_0,d}\ C^{-1}_{n,d,d} \ \Delta \mu_{n,1,1,d}$ 
  \State $\Delta \mu^TC^{-1}B'_{n,w_0,h_0,c_0} \gets  \sum_{d}^{} \Delta \mu_{n,1,1,d}\ C^{-1}_{n,d,d} \ B'_{w_0,h_0,c_0,d}$ 
  \State $\Delta \mu^TC^{-1}\Delta \mu_{n,1,1,1} \gets  \sum_{d}^{} \Delta \mu_{n,1,1,d}\ C^{-1}_{n,d,d} \ \Delta \mu_{n,1,1,d}$ 
  \State $\Gamma \gets  e^{-\frac{1}{2}(B'^TC^{-1}B'+ B'^TC^{-1}\Delta \mu + \Delta \mu^TC^{-1}B' + \Delta \mu^TC^{-1}\Delta \mu)} \cdot I$ \Comment Compute the Gaussian contributions
  \State $P_{n, w_0, h_0, c_0,d} \gets \left \lfloor \mu \right \rfloor_{n,1,1,1,d} + B'$ \Comment Compute the voxel positions that the Gaussians will impact
  \State $Valid_{n,w_0,h_0,c_0} \gets (P_{n,w_0,h_0,c_0,0} \geq 0) \land (P_{n,w_0,h_0,c_0,0} < w) \land (P_{n,w_0,h_0,c_0,1} \geq 0) \land (P_{n,w_0,h_0,c_0,1} < h) \land (P_{n,w_0,h_0,c_0,2} \geq 0) \land (P_{n,w_0,h_0,c_0,2} < c)$ \Comment Get valid indices that are within the volume
  \State $V \gets \texttt{scatter\_add}(V, P[Valid], \Gamma[Valid])$ \Comment Accumulate the contributions at the valid indices in parallel
  \State \Return $V$ \Comment Reconstructed CT volume
  \label{fast}
  \end{algorithmic}
\end{algorithm*}

\section{Adaptive Densification}
\label{densify}
Inspired by the adaptive density control utilized in 3D Gaussian Splatting~\cite{3DGS}, we implement this mechanism for our DGR, aiming to densify the Gaussians within the discretized 3D volume, employing techniques including cloning, splitting, and pruning. The densification strategy is periodically applied throughout the optimization process, striking a balance between the number of Gaussians and the reconstruction quality. 

We clone the Gaussian in under-reconstructed regions into two Gaussians. Specifically, the original and cloned Gaussians share identical positions $\mu$ and covariances $\Sigma$, but their intensities are halved to preserve the total intensity. This process helps to capture the fine details in the under-reconstructed regions. While the gradients of the original Gaussians remain unchanged, the gradients of the cloned Gaussians are set to zero. 

Conversely, in over-reconstructed regions, we split the Gaussian into two smaller Gaussians, keeping their scale $\sigma$ in proportion to the original one. The positions $\mu$ of these new Gaussians are derived from the probability density function (PDF) of the original Gaussian.

Furthermore, we implement a pruning mechanism to eliminate Gaussians that either possess near-zero gradient magnitudes or exhibit a scale $\sigma$ exceeding three times the local box dimension. For these Gaussians, we simply remove them from the reconstruction volume to reduce the computational cost and prevent overfitting. More details of the algorithm are provided in Algorithm \textcolor{iccvblue}{2} and in our code.

\begin{algorithm*}
  
  \caption{Adaptive Densification}
  
  \begin{algorithmic}
  \Require $\mu \in \mathbb{R}^{n \times d}$: mean of Gaussians
  \Require $\sigma \in \mathbb{R}^{n \times 1}$: standard deviation of Gaussians
  \Require $I \in \mathbb{R}^{n \times 1}$: intensity of Gaussians
  \Require $n_{max}$: Maximum allowed quantity of Gaussians (500K by default)
  \Require $\tau$: Minimum gradient value (2e-4 by default)
  \Require $\theta$: Threshold determining Gaussian classification as small or large (0.005 of body diagonal length by default)
  \Require $size$: Box size of each Gaussian ($17$ by default)
  \For{$iteration \gets 100\ \textbf{to}\ max\_iter\ \textbf{step}\ 100$}
  \State $avg\_grad \gets \mu.grad / iteration$
  \State $mask_{clone} \gets (avg\_grad \ge \tau) \ \land (\sigma \le \theta)$
  \State $available\_gaussians \gets n_{max} - n$ \Comment Ensure the total number of Gaussians does not exceed the limit
  \State $n_{clone} \gets \min(available\_gaussians, \sum_{mask_{clone}})$
  \State $sorted\_indices \gets \texttt{argsort}(avg\_grad[mask_{clone}])$
  \State $mask_{clone} \gets mask_{clone} \ \land \ \texttt{top\_k}(sorted\_indices, n_{clone})$ \Comment Select top $n_{clone}$ Gaussians to clone
  \State $I[mask_{clone}] \gets I[mask_{clone}] / 2$ \Comment Halve the intensity of the cloned Gaussians to maintain the total intensity
  \State $\mu_{clone} \gets no\_grad(\mu[mask_{clone}]), \ \sigma_{clone} \gets no\_grad(\sigma[mask_{clone}]), \ I_{clone} \gets no\_grad(I[mask_{clone}])$
  \State $\mu \gets \mu \cup \mu_{clone}, \ \sigma \gets \sigma \cup \sigma_{clone}, \ I \gets I \cup I_{clone}, n \gets n + n_{clone}$
  \State $mask_{split} \gets (avg\_grad \ge \tau) \ \land (\sigma > \theta)$
  \State $available\_gaussians \gets n_{max} - n$
  \State $n_{split} \gets \min(available\_gaussians, \sum_{mask_{split}})$
  \State $sorted\_indices \gets \texttt{argsort}(avg\_grad[mask_{split}])$
  \State $mask_{split} \gets mask_{split} \ \land \ \texttt{top\_k}(sorted\_indices, n_{split})$ \Comment Select top $n_{split}$ Gaussians to split
  \State $\mu_{new} \gets PDF(\mu[mask_{split}], \sigma[mask_{split}], 2)$ \Comment Initialize new Gaussians by PDF sampling
  \State $\sigma_{new} \gets \sigma[mask_{split}] / \sqrt[3]{2} , I_{new} \gets I[mask_{split}]$ \Comment Divide the standard deviation by $\sqrt[3]{2} $ to maintain the total volume
  \State $\mu \gets \mu \cup \mu_{new} - \mu_{split}, \ \sigma \gets \sigma \cup \sigma_{new} - \sigma_{split}, \ I \gets I \cup I_{new} - I_{split}, n \gets n + n_{split}$

  \State $mask_{prune} \gets (avg\_grad) \le \tau \ \lor (\sigma > 3\times size)$
  \State $\mu \gets \mu[ \neg mask_{prune}], \ \sigma \gets \sigma[ \neg mask_{prune}], \ I \gets I[ \neg mask_{prune}]$
  \EndFor
  \State \Return $\mu, \sigma, I$ 
  \label{densification}
  \end{algorithmic}
\end{algorithm*}

\section{Visualization}
\label{sec:vis}
\vspace{1.9mm}
Figures \ref{fig:svct}, \ref{fig:lact}, and \ref{fig:iter} visually depict the reconstruction process across various iterations, with the corresponding iteration number indicated for each image. For comparative purposes, Gaussians for Sparse-View CT are initialized as detailed in the main text, whereas those for Limited-Angle CT are initialized randomly. As the iterations progress, the reconstructed volumes consistently show steady improvement, demonstrating the effectiveness of DGR.

\end{document}